# Recent progress in the physical principles of dynamic ground self-righting

Chen Li, Department of Mechanical Engineering, Johns Hopkins University

**Synopsis**

Animals and robots must self-right on the ground after overturning. Biology research described various strategies and motor patterns in many species. Robotics research devised many strategies. However, we do not well understand how the physical principles of how the need to generate mechanical energy to overcome the potential energy barrier governs behavioral strategies and 3-D body rotations given the morphology. Here I review progress on this which I led studying cockroaches self-righting on level, flat, solid, low-friction ground, by integrating biology experiments, robotic modeling, and physics modeling. Animal experiments using three species (Madagascar hissing, American, and discoid cockroaches) found that ground self-righting is strenuous and often requires multiple attempts to succeed. Two species (American and discoid cockroaches) often self-right dynamically, using kinetic energy to overcome the barrier. All three species use and often stochastically transition across diverse strategies. In these strategies, propelling motions are often accompanied by perturbing motions. All three species often display complex yet stereotyped body rotation. They all roll more in successful attempts than in failed ones, which lowers the barrier, as revealed by a simplistic 3-D potential energy landscape of a rigid body self-righting. Experiments of an initial robot self-righting via rotation about a fixed axis revealed that, the longer and faster appendages push, the more mechanical energy can be gained to overcome the barrier. However, the cockroaches rarely achieve this. To further understand the physical principles of strenuous ground self-righting, we focused on the discoid cockroach's leg-assisted winged self-righting. In this strategy, wings propel against the ground to pitch the body up but are unable to overcome the highest pitch barrier. Meanwhile, legs flail in the air to perturb the body sideways to self-right via rolling. Experiments using a refined robot and an evolving 3-D potential energy landscape revealed that, although wing propelling cannot generate sufficient kinetic energy to overcome the highest pitch barrier, it reduces the barrier to allow small kinetic energy from the perturbing legs to probabilistically overcome the barrier to self-right via rolling. Thus, only by combining propelling and perturbing can self-righting be achieved, when it is so strenuous; this physical constraint leads to the stereotyped body rotation. Finally, multi-body dynamics simulation and template modeling revealed that the animal's substantial randomness in wing and leg motions help it by chance to find good coordination, which accumulates more mechanical energy to overcome the barrier, thus increasing the likelihood of self-righting.

**Introduction**

Righting oneself from being upside down on the ground is a prevalent locomotor maneuver that animals must make to survive. Even on level, flat, solid ground with high friction, locomotion can result in overturning. On uneven [1], sloped, or slippery surfaces, overturning is even more likely. Falling during jumping, climbing, or flying, as well as fighting [2] and courtship [3], can also lead to overturning. Once overturned, animals must self-right promptly to avoid predation, starvation, and dehydration. Many animals need to self-right even simply after sleep. Similarly, mobile robots can flip over during a diversity of locomotor tasks [4,5]. How likely and quickly animals and robots can self-right on the ground is important for their survival or continuing and timely operation.

Ground self-righting behavior and motor patterns have been extensively studied in insects, such as cockroaches [6–10], beetles [11–16], stick insects [17], locusts [18], stink bugs [19], lanternflies [20], and springtails [21], as well as in other animals such as crustaceans [22,23], mollusks [24–26], toads [27], lizards [28], turtles [29–35], birds [36], and mammals [37,38]. Many biological ground self-righting strategies have been described, including: (1) using appendages (e.g., legs, wings, tail, antennae, ventral tube) and neck/head to grasp, pivot, push, pull, or shake [6,12,23–27,29–33,13,34–38,14,15,18–22], (2) deforming the body [7,23,28,37], (3) having a body shape and center of mass that makes an upside-down orientation unstable [31,39], and (4) jumping with elastic energy storage and release then falling into an upright orientation by chance [11,12,16]. Different types of appendages and body deformation are often used together [15,18,37,38,21,23,26,29–31,33,36]. These diverse strategies lead to self-righting via diverse body rotations, including pitching, rolling, and diagonal rotations with both pitching and rolling [6,7,22–25,27,30–34,12,37,13–15,18–21]. Many species use multiple strategies and transition among them to self-right [7,12,14,15,19,20,29,37]. For robots, a diversity of terrestrial self-righting strategies have also been developed, including all four categories above, as well as having a symmetric body design without an upright orientation (for a brief review, see [4,5]).

Given these rich descriptions of biological strategies and motor patterns as well as plentiful development of robot strategies, we know relatively little about the physical principles of how the fundamental need to generate mechanical energy (kinetic energy and potential energy) to overcome the potential energy barrier to self-right on the ground governs behavioral strategies and body rotations given the morphology.

Here, I review the major findings from recent studies [4,5,40–43] that I led to begin to fill this knowledge gap, focusing on cockroaches and their robophysical models [44]. Our inquiries began with developing a cockroach-inspired robot, which relies on opening its wings to self-right dynamically (Fig. 1) [4,5]. Inspired by a 2-D potential energy landscape model for turtle ground self-righting (Fig. 2), we tested this initial robot to reveal the physical principles of dynamic ground self-righting via a fixed-axis body rotation to overcome the potential energy barrier (Fig. 3) [4,5]. To test whether cockroaches use such simple rotations, we performed animal experiments to quantify how three species of cockroaches use and transition across various self-righting strategies (Figs. 4-6) [40]. We developed a simplistic 3-D potential energy landscape model and measured the animals' often complex yet stereotyped 3-D body rotations (Figs. 7-8) to explain why each species rolls more during successful attempts than failed ones [40]. More interesting questions arose from our animal observations—that ground self-righting is often dynamic (using kinetic energy to overcome the potential energy barrier) yet strenuous, the body rotation is stereotyped, propelling motions are accompanied by perturbing motions, and there is substantial randomness in these motions. To better understand the physical principles governing these, we then focused on the discoid cockroach's strenuous, leg-assisted winged self-righting as a model system (Fig. 9). By combining robophysical modeling and evolving potential energy landscape modeling with a refined robot (Figs. 10, 11) [41], multi-body dynamics simulation (Fig. 12) [42], and template modeling (Fig. 13) [43], we elucidated the physical constraints that lead to stereotyped body rotation, and how and why the cooperation, coordination, and substantial randomness in the motions of propelling wings and perturbing legs contribute to successful self-righting. This brief review focuses on threading together the major approaches and findings of these studies. For a deeper dive into the focus, motivation, methods, results, implications, and limitations of each study, please refer to the original research papers.

Before reviewing our work, we note that the physical principles of ground self-righting dominated by ground reaction forces differ from aerial and underwater self-righting dominated by different forces. When aerodynamic forces are negligible, aerial self-righting is governed by the conservation of angular momentum, and animals can rotate their appendages to induce counter-rotations of the body to self-right (e.g., [45]). When aerodynamic forces dominate, animals can control their aerodynamic surfaces to generate rotating torques to self-right in the air (e.g., [46]). The physical principles of aerial self-righting have been well understood; for a comprehensive review, see [47]. Similarly, underwater self-righting that predominantly uses hydrodynamic forces, with little substrate interaction (e.g. [48]), should have physical principles more similar to aerial self-righting in the limit when aerodynamic forces dominate (e.g., [46]). However, underwater ground self-righting that largely relies on interaction with the bottom substrates [22–26] should be governed by similar principles as ground self-righting, except that the larger hydrostatic and hydrodynamic forces need to be taken into account (whereas the buoyant and drag forces in the air is negligible during ground self-righting).

**How it all started**

In an earlier study of how insects traverse cluttered, grass-like beam obstacles, we discovered that the discoid cockroach's rounded body shape helps it roll into narrow gaps between obstacles to traverse (Fig. 1A) [49]. A cockroach-inspired rounded shell enabled a legged robot to roll its body into obstacle gaps to traverse similar cluttered obstacles (Fig. 1B). However, the robot sometimes over-rolls and flips over, and it gets stuck (Fig. 1B), whereas the cockroach sometimes flips over, too, but can recover (Fig. 1A) by using its wings to push against the ground (Fig. 1C). This led us to develop the robot's rounded shell into two actuated wings that push against the ground to self-right via body pitching (Fig. 1D) [4,5]. In this process, I became interested in understanding the physical principles of ground self-righting.

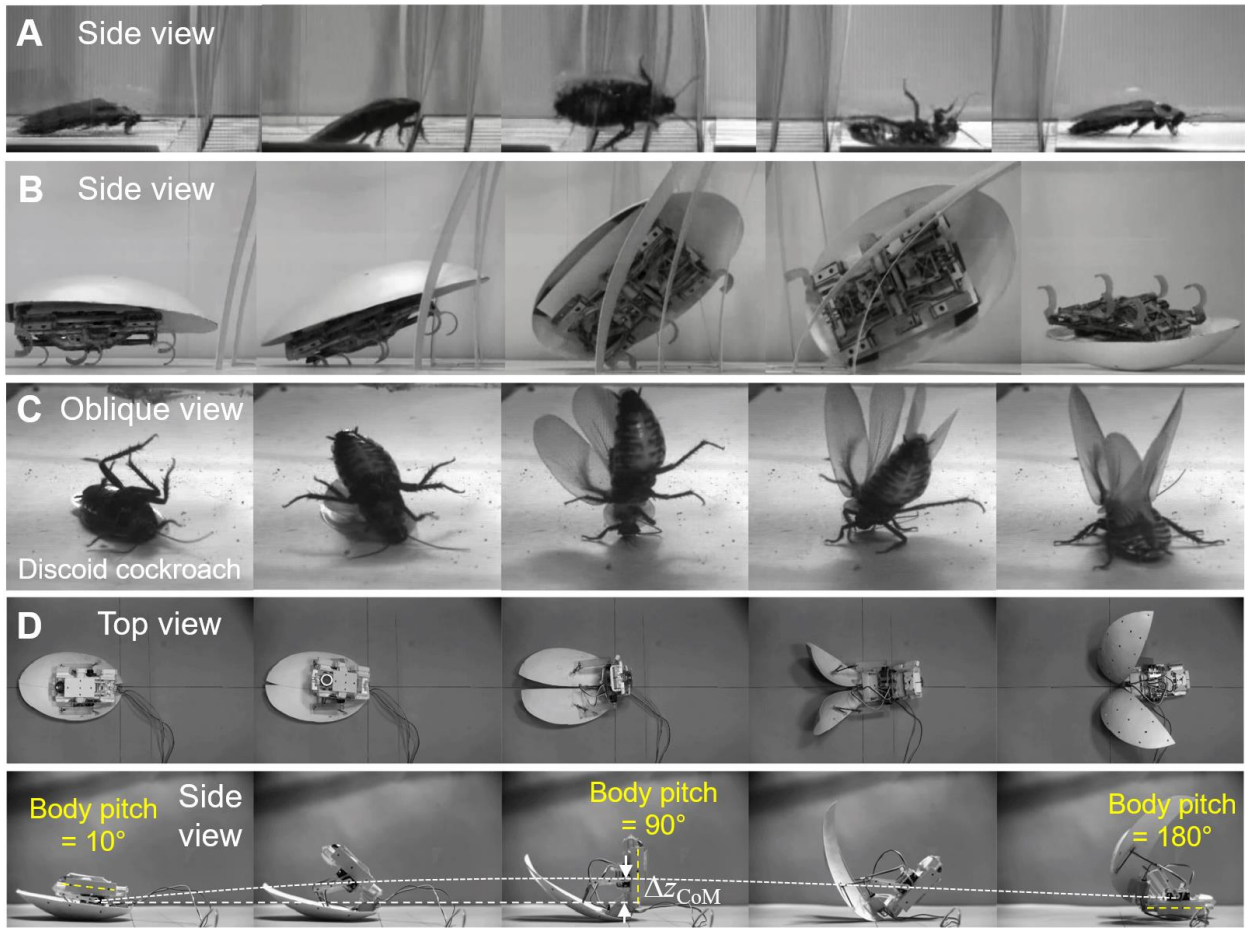

**Fig. 1. Study of cockroach and robot traversing cluttered obstacles led to study of cockroach and robot ground self-righting.** Adapted from [5]. **(A)** A discoid cockroach traverses cluttered, grass-like beam obstacles, during which its wings are folded against the body as a rounded ellipsoidal "shell" to facilitate body rolling into obstacle gaps. It flips over upon exiting the obstacles and quickly rights itself. **(B)** A small six-legged robot uses a cockroach-inspired rounded shell to traverse cluttered obstacles. However, when it over-rolls and flips over, it cannot self-right. **(C)** The cockroach self-rights by opening and pushing its wings against the ground to pitch and roll its body. **(D)** The robot with the rounded shell cut into two wings, which open to push against the ground, to self-right via pure body pitching. Change in body pitch is shown by yellow dashed lines. White dashed curve shows that center of mass (CoM) height increases by $\Delta z_{CoM}$ (defined by arrows) from an upside-down orientation (left most) to the highest CoM orientation (middle), then decreases. Potential energy barrier is $\Delta E_{potential} = mg\Delta z_{CoM}$, where $m$ is total body mass and $g$ is gravitational acceleration. Adapted from [5].

**Inspiration from modeling of turtle ground self-righting**

To self-right on the ground, an animal or robot must change its body orientation from upside down to upright (e.g., change body pitch from ~ 0° to 180°, Fig. 1D, bottom; change body roll from 0° to 180°, Fig. 2A, bottom), which requires overcoming a gravitational potential energy barrier (Fig. 1D, Fig. 2A). A previous study used a 2-D potential energy landscape to model turtles self-righting via body rolling in the transverse plane (Fig. 2) [31]. It well explained why turtles of highly domed shells with a low (or even diminished) barrier can simply use passive body rolling complemented by leg and neck motions to self-right (Fig. 2B), whereas turtles with flatter shells leading to a higher barrier must more vigorously use their legs and neck to push against the ground to self-right (Fig. 2A). A similar potential energy landscape framework has also been established for motion planning of robots using an appendage to self-right quasi-statically on sloped planar surfaces in the sagittal plane [50].

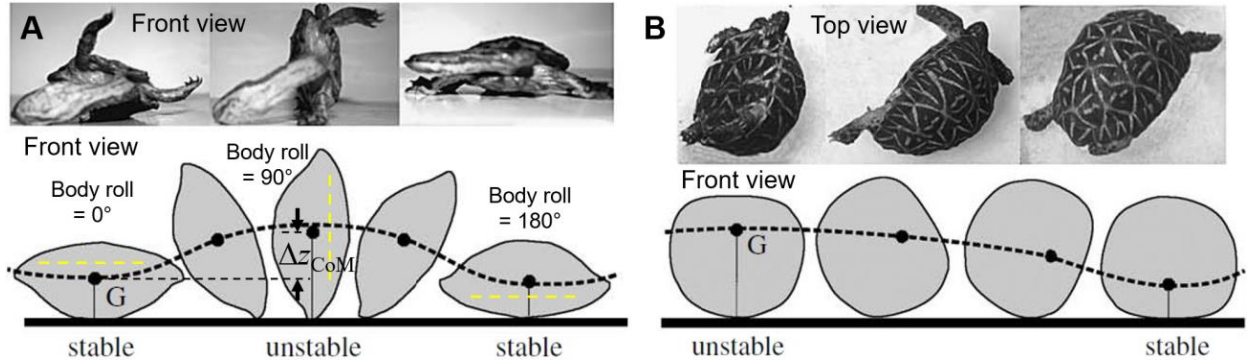

**Fig. 2. Potential energy landscape for turtle ground self-righting in 2D via body rolling.** (**A**) Potential energy barrier, $\Delta E_{\text{potential}} = mg\Delta z_{\text{CoM}}$, is high for turtles with a flatter shell. Black dashed curve shows that center of mass (CoM) height increases by $\Delta z_{\text{CoM}}$ (defined by black arrows) from an upside-down orientation (left most) to the highest CoM orientation (middle), then decreases. Potential energy barrier is $\Delta E_{\text{potential}} = mg\Delta z_{\text{CoM}}$, where $m$ is total body mass and $g$ is gravitational acceleration. Change in body roll is shown by yellow dashed lines. (**B**) Potential energy barrier to self-right diminishes for turtles with a highly domed shell and low center of mass when upright. Adapted from [31].

**Principles of ground self-righting with single-axis body rotations**

To understand the physical principles of ground self-righting, we first systematically studied our initial cockroach-inspired robot (Fig. 1D) [4,5]. Because its body is longest in the longitudinal direction, when the robot self-rights via body pitching, it has to overcome the largest potential energy barrier. We varied the wing opening amplitude $\Theta_{\text{wing}}$ and speed $\omega_{\text{wing}}$ (how much and how fast the wings open) to test how they affect the robot's ability to overcome this largest barrier (Fig. 3). The more and faster the wings open, the more likely the robot is to self-right (Fig. 3A, B), and the shorter the time it takes (Fig. 3A, C).

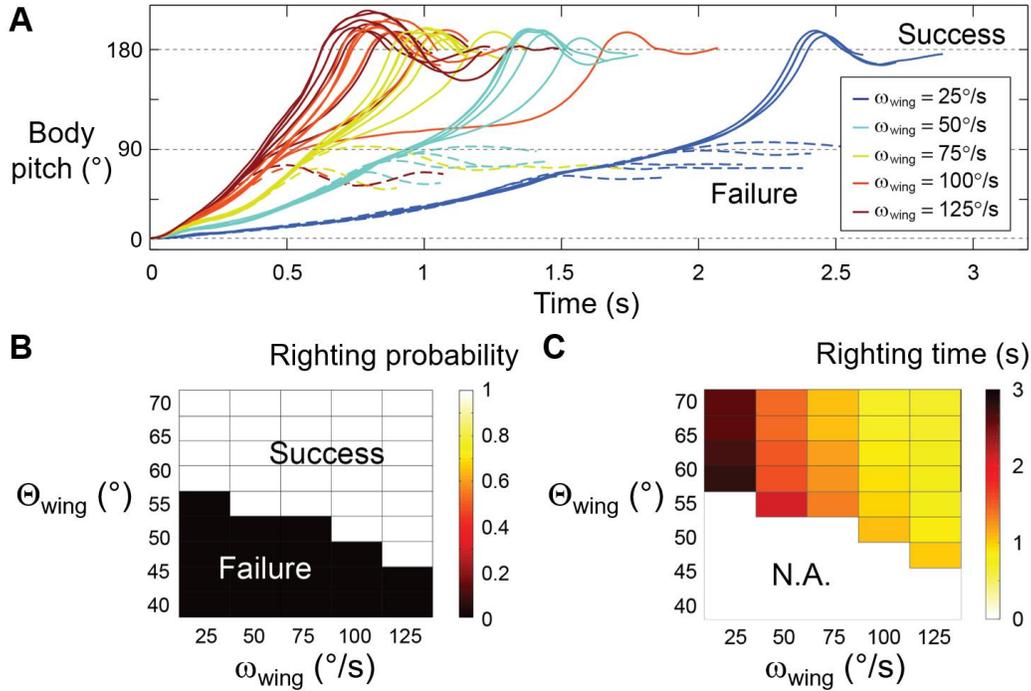

**Fig. 3. How appendage pushing magnitude and speed affect self-righting via a simple rotation about a fixed axis, from experiments using the initial robot** (Fig. 1D). **(A)** Body pitch as a function of time, for a wide range of wing opening amplitude $\Theta_{wing}$ and speed $\omega_{wing}$. Each curve of the same color shows one of the three trials that uses the same $\omega_{wing}$ (value in legend) but a different $\Theta_{wing}$. Solid and dashed curves show successful and failed trials, respectively. **(B, C)** Righting probability and average righting time as a function of $\Theta_{wing}$ and $\omega_{wing}$. Adapted from [5].

This initial robot experiment revealed intuitive physical principles of ground self-righting using appendages to generate a simple body rotation about a fixed axis: The longer and faster the appendages propel against the ground, the more mechanical energy can be gained to overcome the barrier to self-right on the ground [4,5]. However, do animals always use such simple body rotations to self-right? If not, what are the physical principles that govern more complex self-righting via 3-D body rotations? Can we understand these principles using a potential energy landscape approach?

**Ground self-righting may require multiple attempts**

To explore these questions, we studied the Madagascar hissing, American, and discoid cockroaches self-righting on a level, flat, solid, low friction surface [40]. All three species always self-right if given sufficient time (near 100% probability within 30 seconds, Fig. 4A, white). However, although in some trials the animals can self-right upon the first attempt, in other trials they struggled, requiring multiple attempts to self-right (Fig. 4B; Fig. 4A, gray). Although they can self-right within a short time (~ 1 second) if the first attempt is successful, when multiple attempts are needed, self-righting can take much longer (up to ~ 10 seconds).

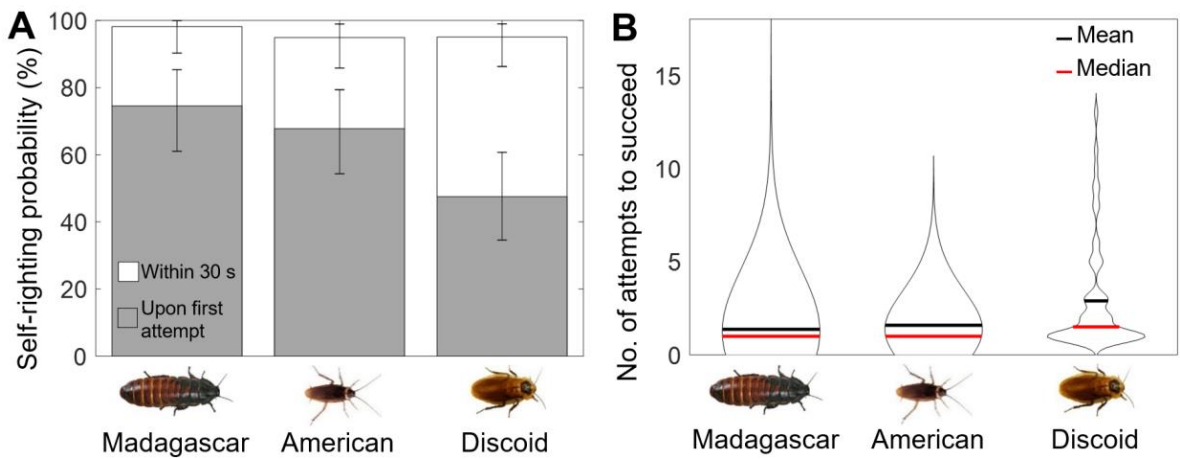

**Fig. 4. Self-righting performance.** (**A**) Self-righting probability within 30 seconds (white) and on the first attempt (gray). Error bars represent 95% confidence intervals. (**B**) Violin plots of the number of attempts required to achieve self-righting. Width of graph shows the relative frequency of the data along the *y*-axis. Black and red lines show mean and median for each species. Adapted from [40].

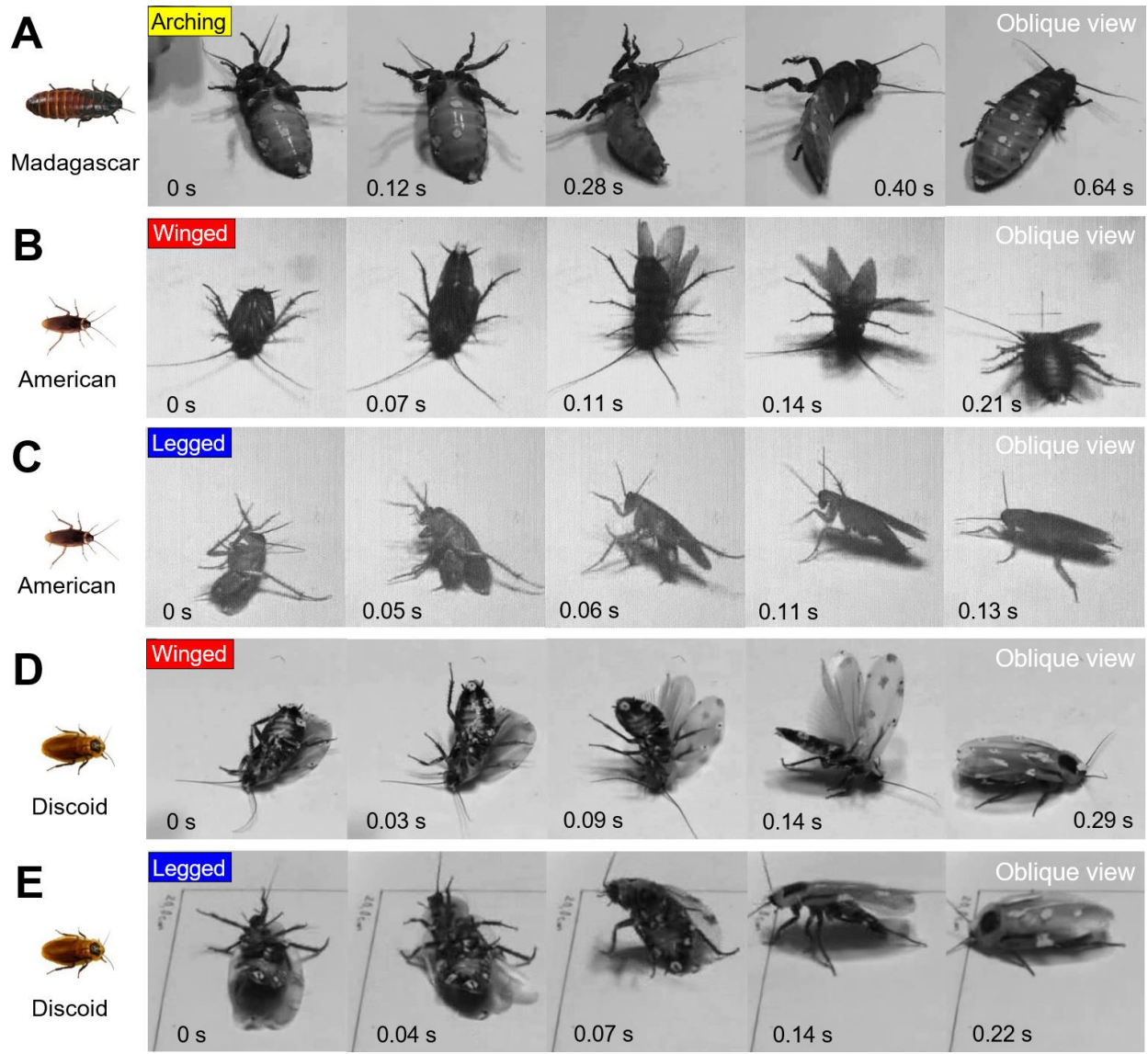

**Fig. 5. Self-righting strategies that lead to success. (A)** Madagascar hissing cockroach using body arching to self-right. **(B, C)** American cockroach using wings or legs to self-right. **(D, E)** Discoid cockroach using wings or legs to self-right. Adapted from [40].

**Cockroaches use and transition across diverse strategies and often self-right dynamically**

All three species attempt to use more than one strategy to self-right (e.g., Fig. 5) and often transition across them (Fig. 6) [40]. The Madagascar hissing cockroach often hyperextends its body into an arch to roll onto one side, followed by leg scrubbing against the ground (Fig. 5A), which almost always leads to successful self-righting eventually (Fig. 6A, yellow). Occasionally, it twists the body in an attempt to find objects the legs can grasp onto, but this never leads to successful self-righting (Fig. 6A, green). Both the American and discoid cockroaches can use two strategies (Fig. 6, B, C) that can lead to successful self-righting, with wings (Fig. 5B, D) or legs (Fig. 5C, E) as the main propelling appendages to push against the ground, respectively. Notably, both the American and discoid cockroaches often self-right dynamically, by gaining sufficient pitch and roll kinetic energy from wings or legs pushing against the ground to overcome the potential energy barrier. The American cockroach also very occasionally flaps its wings, which always fails in righting itself (Fig. 6B, cyan). All species sometimes enter quiescence without apparent movement (Fig. 6A-C, white oval).

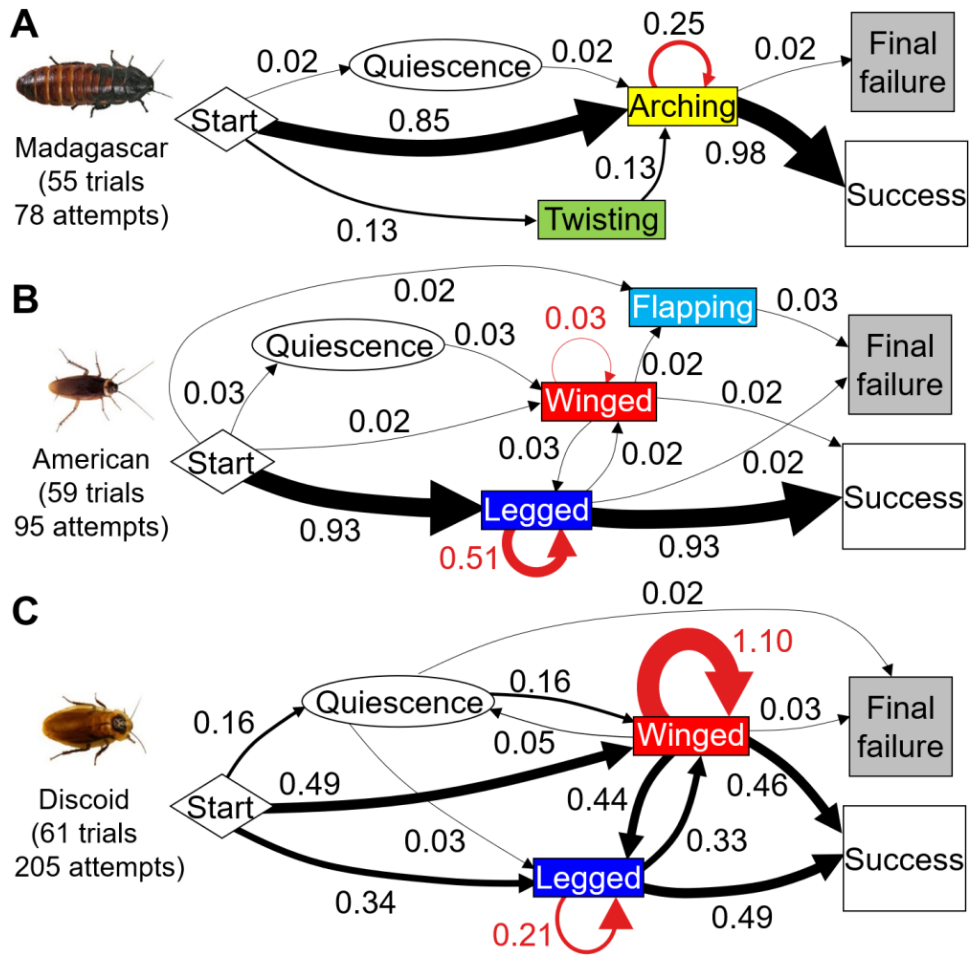

**Fig. 6. Self-righting locomotor transition ethograms.** (**A**) Madagascar hissing cockroach. (**B**) American cockroach. (**C**) Discoid cockroach. Arrow widths are proportional to transition relative frequencies, with values shown by numbers. Relative frequency is defined as the ratio of the number of occurrences of each transition to the total number of trials for each species. The sum of relative frequencies out of each node equals that into the node, except for start with a sum of 1 going out, and success and final failure with a sum of 1 into both together. (Final failure occurs in some trials, when the animal is able to self-right before the trial conclude at 30 seconds; in other trials, the animal can fail multiple attempts but eventually succeeds in self-righting.) Red arrows and numbers show probabilities of self-transitions (into the same node) and represent the average number of times of continuing the same strategy during each trial. A self-transition probability greater than one means that on average it occurred more than once for each trial. Only the strategies that can lead to successful self-righting on the level, flat, solid, low-friction ground are shown in Fig. 5. Adapted from [40].

**Propelling motions are accompanied by perturbing motions with substantial randomness**

The winged and legged strategies (Fig. 5C-E) often involve using more than a single type of appendages or even deforming the body [40]. During winged self-righting, the discoid cockroach also flexes and twists the abdomen, flails the legs, and/or scrapes the legs against the ground (Fig. 5D). During legged self-righting, both the American and discoid cockroaches also flex and/or twist the abdomen (Fig. 5C, E). In other words, the primary propelling appendages (wings in the winged strategy; legs in the legged strategy) are accompanied by assisting motions by other appendages or body deformation, which provides additional perturbations.

In addition, both the propelling and perturbing motions are quite erratic, with large variations in the direction, magnitude, frequency, and coordination (phase offset between various motions) from attempt to attempt [40]. The randomness in these motions is much larger than that in the highly rhythmic leg oscillations during walking [51] and running [52]. Previous neurophysiological studies also showed that leg activation patterns during ground self-righting are more random than during walking [9,10].

**Stochasticity and stereotypy of use of self-righting strategy**

In any single trial, it is stochastic which strategy an animal will use or transition to, or if it will continue using the same strategy, over each attempt. However, averaging over large numbers of trials (see sample size in Fig. 6), the behavioral pattern of each species using strategies is stereotyped (Fig. 6) [40]. Here, behavioral stereotypy means that the actual observed behavior is a small fraction of a large number of possibilities (see [53,54]). Specifically, both the Madagascar hissing and American cockroaches predominantly rely on a single strategy (body arching and legged strategy, respectively) to self-right (Fig. 6A, B). By contrast, the discoid cockroach has a more balanced use of two strategies (winged and legged) (Fig. 6C).

**Simplistic 3-D potential energy landscape reveals that more body rolling is advantageous**

We observed that the cockroaches' body rotations during ground self-righting are rarely about a fixed axis in the pitch-roll space, but often complex, non-planar (e.g., Fig. 5B-E). Thus, to understand the physical principles, we needed to expand the potential energy landscape approach into three dimensions. As a first step, we developed a simplistic 3-D potential energy landscape [40] (Fig. 7). By approximating each cockroach species' body shape with an ellipsoid of similar dimensions, we calculated its potential energy as a function of body pitch and roll using Euler angles (Fig. 7B). These animals' body length is greater than body width, which is greater than body height. Thus, self-righting by pure body pitching overcomes the highest potential energy barrier (Fig. 7A; Fig. 7B, cyan), and self-righting by pure body rolling overcomes the lowest barrier (Fig. 7D; Fig. 7B, magenta). Self-righting using body rotations with simultaneous pitching and rolling overcomes an intermediate barrier (e.g., Fig. 7C; Fig. 7B, green, yellow), and the more it rolls, the lower the barrier becomes (Fig. 7B, green vs. yellow). Pure body yawing without pitching or rolling cannot raise the center of mass.

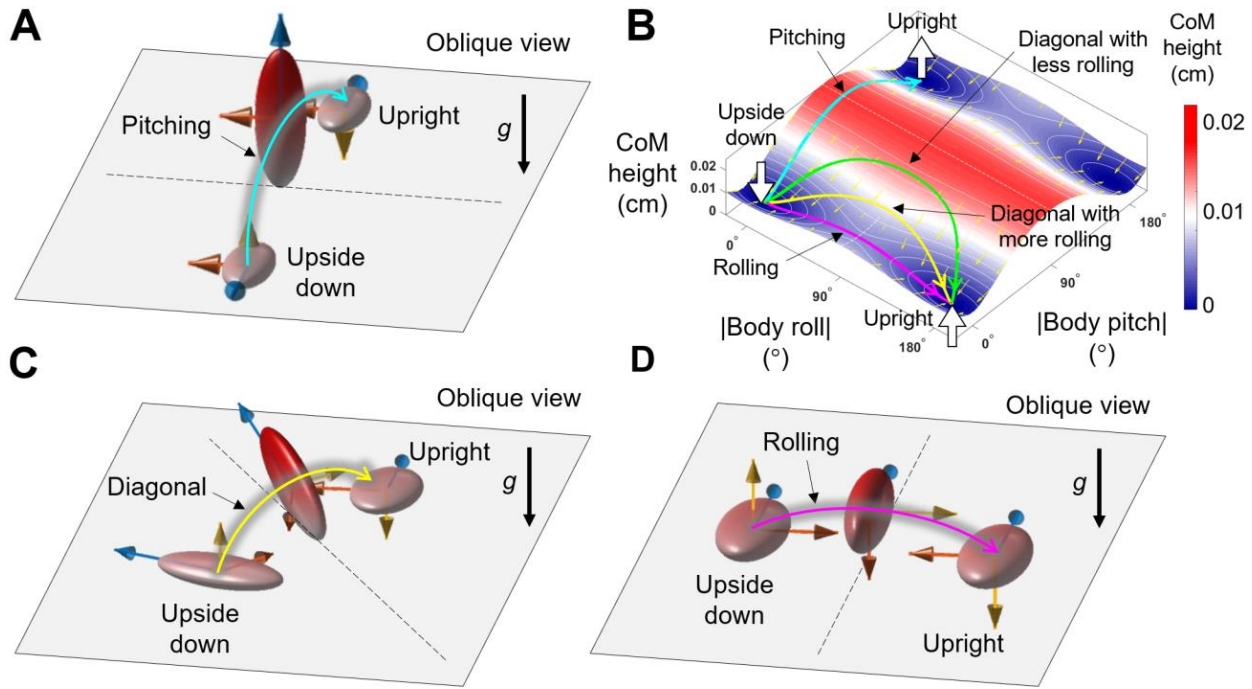

**Fig. 7. Simplistic 3-D potential energy landscape of an ellipsoidal rigid body. (A, C, D)** An ellipsoid approximating the animal body in contact with the ground, either pitching (A), rolling (D), or rotating diagonally, with simultaneous pitching and rolling (C). (A-C) all show simple rotations about a fixed axis (dashed line) within the horizontal ground plane. Actual rotation of the animal body may be about a time-varying axis. Red, blue, and yellow arrows on each ellipsoidal body show its three major axes to illustrate body rotation. Vector *g* shows the direction of gravity. **(B)** Potential energy landscape, shown as CoM height as a function of body pitch and roll, using Euler angles with yaw-pitch-roll convention. We use absolute values of body pitch and roll, considering symmetry of the ellipsoid. Downward and upward white arrows indicate an upside-down and upright body orientations, respectively. Cyan, green, yellow, and magenta curves with arrows are representative trajectories for pure pitching, two different diagonal rotations, and pure rolling, each about its own fixed axis in the horizontal plane, to illustrate the decrease of potential energy barrier with more body rolling. White curves on the landscape are iso-height contours. Small yellow arrows on the landscape are gradients. Model results shown are using the discoid cockroach's body dimensions as an example. Adapted from [40].

**Self-righting body rotations can be complex and are stereotyped**

We then examined each species' body pitch and roll at three stages of each attempt (start, highest center of mass (CoM) orientation, and end) on the simplistic 3-D potential energy landscape to assess how it rotates and how this affects the potential energy barrier (Fig. 8) [40]. For the Madagascar hissing cockroach using body arching, body rotation is mainly rolling (Fig. 8A, Fig. 5A), which overcomes the lowest potential energy barrier when successful. For the American cockroach using wings, body rotation is mainly pitching (Fig. 8B, Fig. 5B), which overcomes the highest barrier when successful. For the American cockroach using legs, body rotation is mainly rolling with a small amount of pitching (Fig. 8C, Fig. 5C), which overcomes nearly the lowest potential energy barrier when successful. By contrast, for the discoid cockroach using both wings and legs, body rotation has both large pitching and large rolling (Fig. 8D, E, Fig. 5D, E), which overcomes an intermediate potential energy barrier if successful. In addition, each species' body rotation is stereotyped, reaching a similar orientation when CoM is highest, whether the attempt is successful or not (Fig. 8A-E, small variation of state 2 orientation). This stereotypy suggested that physical constraints strongly confine the body rotation [54]. All three species also often have large body translation and yawing in the horizontal plane from appendage interaction with the level, flat, solid, low friction ground, but this does not contribute to self-righting as it cannot raise the center of mass.

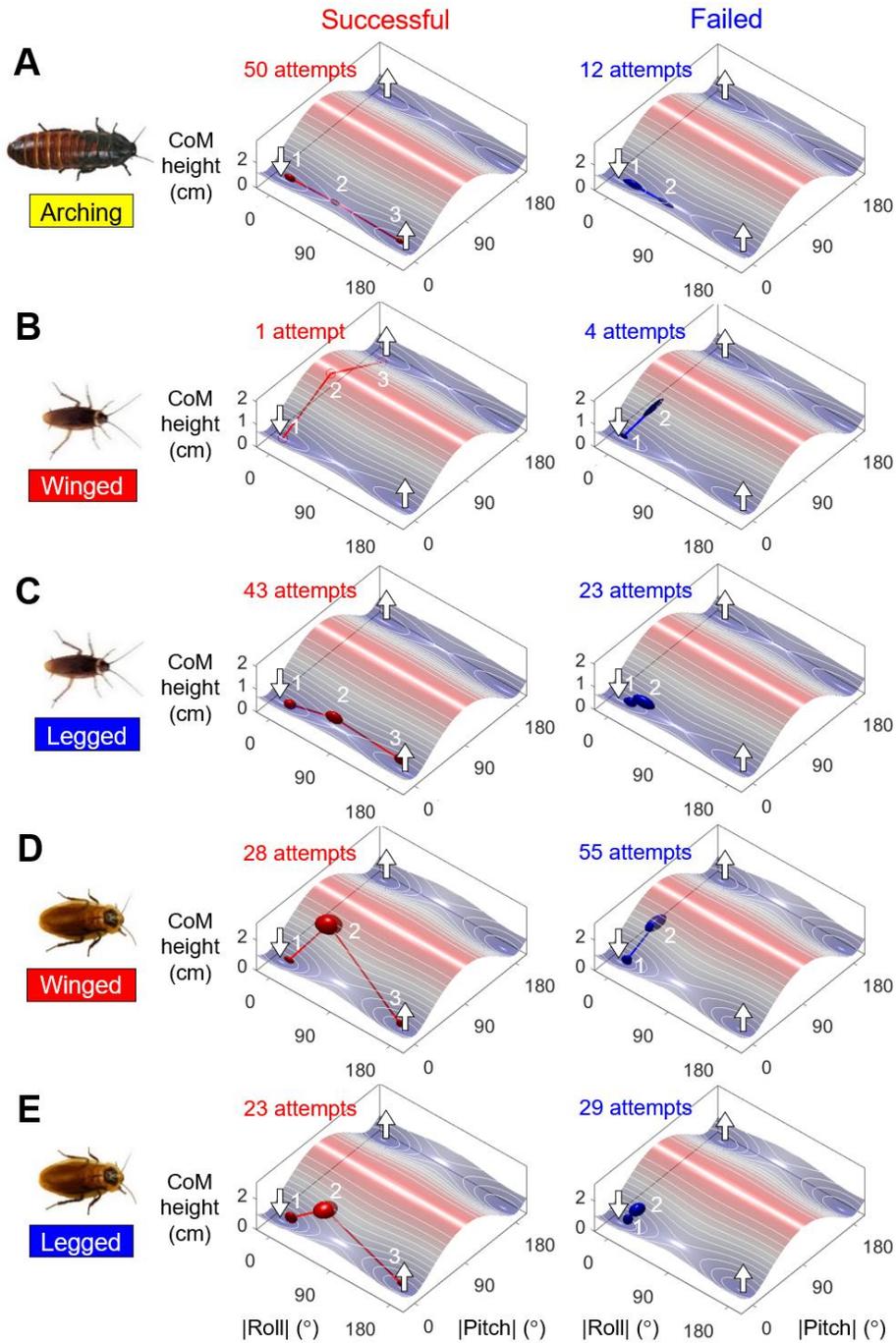

**Fig. 8. State of the body on the potential energy landscape at the start (1), highest CoM position (2), and end (3) of the attempt during successful vs. failed attempts.** (**A**) Madagascar hissing cockroach using body arching. (**B, C**) American cockroach using wings or legs. (**D**) Discoid cockroach using wings or legs. Potential energy landscape is defined in Fig. 5. On each landscape, the ellipsoids show means

(center of ellipsoid) ± 1 s.d. (principal semi-axis lengths of ellipsoid) of body pitch, body roll, and CoM height at each stage of the attempt. For failed attempts (right), the end state (3) is not shown because it nearly overlaps with the start state (1). The number of attempts of each case is shown. Adapted from [40].

**Cockroaches roll more in successful attempts**

All three species roll their body more during successful self-righting attempts than in failed ones (Fig. 8, left vs. right) [40]. The higher body rolling in successful attempts lowers the potential energy barrier (Fig. 7B, C), making it easier to be overcome to achieve self-righting, given the mechanical energy that can be generated. Consistent with this finding, in our experiments using the initial robot (Fig. 1D), we also tested opening the two wings asymmetrically, where we found that with the more the body rolls the more likely the robot is to self-right [4,5].

**Ground self-righting is strenuous for cockroaches**

However, for all three species using the strategies that lead to success, the increase in CoM height from the start to the highest CoM orientation (Fig. 8, from state 1 to state 2) is only slightly larger during successful attempts than in failed ones (Fig. 8, left vs. right) [40]. In other words, only a small difference in how much an animal can raise its CoM determines whether it succeeds or fails in overcoming the potential energy barrier. This, together with the observation that they often require multiple attempts to succeed (Fig. 4B), provide evidence that ground self-righting is strenuous for these cockroaches: they can barely do enough work during each attempt to overcome the potential energy barrier [55]. Previous force measurements also support this notion: for the discoid cockroach to self-right using legs, a single hind leg needs to generate a ground reaction force as large as eight times that during high speed running [6].

**Strenuous, leg-assisted winged self-righting as a model system to study new questions**

Our animal observations quantified stochastic yet stereotyped behavioral transitions and often complex yet stereotyped 3-D body rotations, and our simplistic model explained why successful attempts have more body rolling. However, some new questions arose. First, why are the body rotations stereotyped, even though any arbitrary 3-D rotations in the body pitch and roll space are in principle possible? Second, are the perturbing motions that accompany the motions of propelling appendages useful? Third, is the substantial randomness in the motions beneficial?

To further understand these, we performed three additional studies [41–43], focusing on a model system—the discoid cockroach's strenuous, leg-assisted winged self-righting [40] (Fig. 9). In this strategy, the overturned animal always first opens and pushes its wings against the ground to pitch up the body (Fig. 9A, B, blue). Because the two wings open together, the center of mass falls within a triangular base of support formed by the two opened wings and head (Fig. 10A, black dashed triangle) [41]. This intermediate state is metastable (i.e., stable provided that it is subjected to only small perturbations [56]). However, wing pushing rarely pitches the animal sufficiently to complete a full somersault (Fig. 9A, B, dashed blue arrow), and the animal often continually attempts wing pushing but fails to self-right (Fig. 9A, B, solid blue arrows). When it eventually succeeds, the animal almost always rolls sideways over one of the wings from the metastable state (Fig. 9A, B, red arrow). Throughout this process, the animal often vigorously flails its legs in the air (Fig. 10A, red dashed curves). The legs also sometimes scrape the ground, the abdomen flexes and twists, and the wings often deform passively under load. All these motions, which have substantial randomness, may result in perturbations in the roll direction.

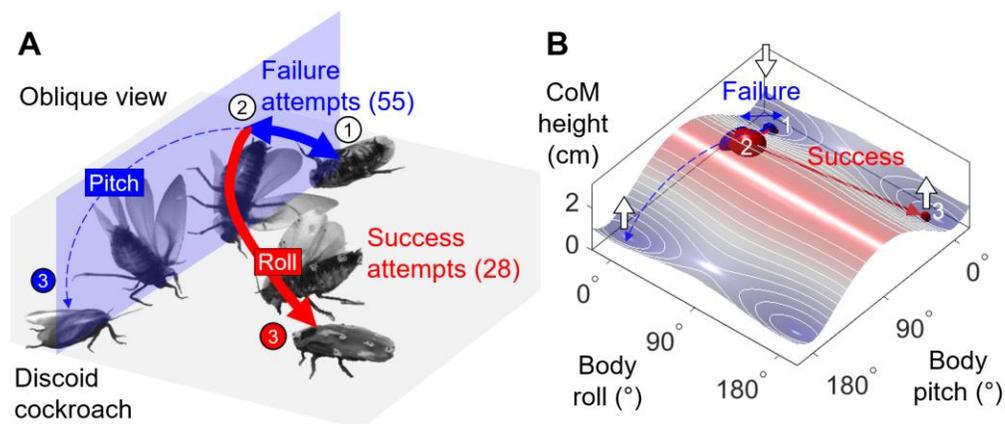

**Fig. 9. Strenuous leg-assisted, winged self-righting of discoid cockroach as a model system. (A)** Representative snapshots of the animal, illustrating body rotations during failed pitching attempts (thick

blue arrows), a successful attempt by pure pitching (thin dashed blue arrow), and a successful attempt by pitching up first then rolling (red arrow). Adapted from [41]. **(B)** Stereotyped body rotation observed in successful (red) and failed (blue) attempts, overlaid on the potential energy landscape. In (A, B), 1, 2, and 3 show upside-down (1), metastable (2), and upright (3) states. This plot is the same as Fig. 8D, left and right merged, except that the axes are flipped to show the three stages' trajectory in a similar view as in Fig. 9A. See Fig. 8 for definition of elements.

Specifically, for this model system, we need to answer the following three questions. First, why is the discoid cockroach's body rotation not mainly rolling, which has the lowest potential energy barrier, but instead first pitching then rolling when successful? Second, is the leg flailing motion that accompanies the wing opening motion useful? Third, is the substantial randomness in the coordination of the wing and leg motions beneficial?

**Refined robophysical model of strenuous leg-assisted winged self-righting**

To address these questions, we first created a refined robot as a robophysical model of strenuous, leg-assisted winged self-righting, following biological observations (Fig. 10A, B) [41]. This robot is not aimed to achieve self-righting; that has already been done by the previous robot (Fig. 1D). Instead, we deliberately designed and controlled this robot to achieve similar *strenuous* self-righting behavior as the discoid cockroach's (Fig. 9A). Like the discoid cockroach (Fig. 10A), the refined robot has a head protruding forward from the body, creating the triangular metastable state (Fig. 10B, black dashed triangle), which the initial robot lacks. In addition, we limited the refined robot's wing opening amplitude so that it cannot self-right via pure body pitching. These makes it a biologically relevant robophysical model for studying this strenuous strategy.

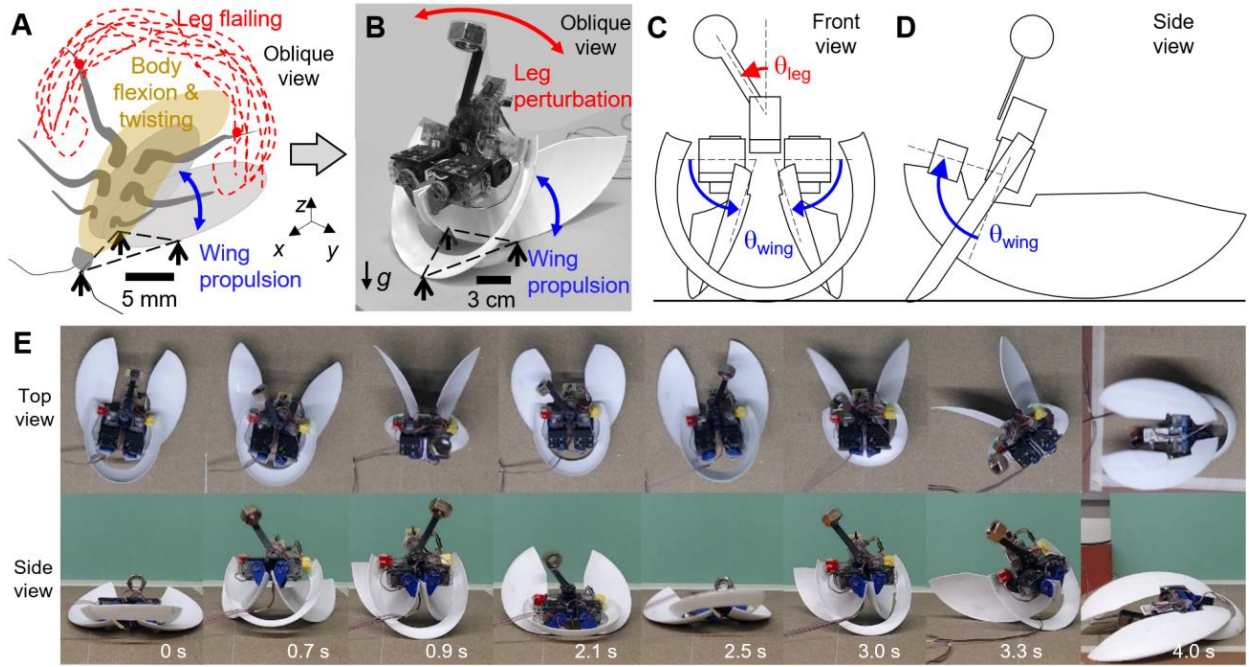

**Fig. 10. Robophysical modeling of strenuous leg-assisted, winged self-righting.** (**A**) Schematic of animal in metastable state. Blue arrows show wing opening (and closing) to propel against the ground. Red dashed curves show vigorous leg flailing. *x-y-z* is lab frame. (**B**) Refined robot in metastable state. In (A, B), dashed black triangle shows base of support, formed by ground contacts of head and two wing wedges. (**C, D**) Front and side view schematics of robot in metastable state, to define leg angle $\theta_{leg}$ (red) and wing angle $\theta_{wing}$ (blue). Both wings rotate simultaneously. Each wing pitches away from the body (D) as well as rolls (E) to open. At any moment during wing opening and closing, wing pitching and rolling always reach the same angle $\theta_{wing}$. (**E**) Representative snapshots of robot self-righting after two attempts. (A-D) are adapted from [41]. (E) is adapted from [42].

Robophysical modeling allows systematic parameter variation to discover physical principles of locomotion involving complex motions and locomotor–environment interactions [44]. For leg-assisted winged self-righting, we need to systematically vary propelling wing motion and perturbing leg motion. To generate wing motion similar to that of the discoid cockroach, the refined robot opens both wings symmetrically, rolling and pitching them about the body by the same angle $\theta_{wing}$ (Fig. 10C, D, blue arrows) to propel against the ground. Because the animal's perturbing motions are highly complex, to simplify parameter variation in the robot, we focused on the more frequent leg flailing. We chose to use a one degree-of-freedom, pendulum-like "leg", which oscillates laterally by the same amplitude to both sides, to generate perturbation (Fig. 10B, red arrow). Besides having similar geometric proportions to the animal (Fig. 10B vs. A), we also verified that the robot's leg actuation generated dynamically similar motion as that of the animal [41].

By opening and closing its wings repeatedly while oscillating the leg, the refined robot generates similar strenuous self-righting attempts, with similar motions as observed in the animal, not being able to self-right by pure body pitching, and often requiring multiple attempts to self-right via body rolling after first pitching up to the metastable state (e.g., Fig. 10E).

**Propelling and perturbing appendages together enable barrier-crossing to self-right**

In our experiments, we used the refined robot to measure the full 3-D body rotation and wing motion during entire self-righting trials, which is challenging to measure for the animals due to the frequent occlusions [40]. This enabled us to reconstruct an accurate, "evolving" potential energy landscape, which changes with wing opening and closing, rather than the simplistic, fixed landscape from a rigid body (Figs. 7-9). We used this evolving potential energy landscape to understand how the propelling motion of the wings and perturbing motion of the leg together allow overcoming the potential energy barrier to self-right (Fig. 11) [41].

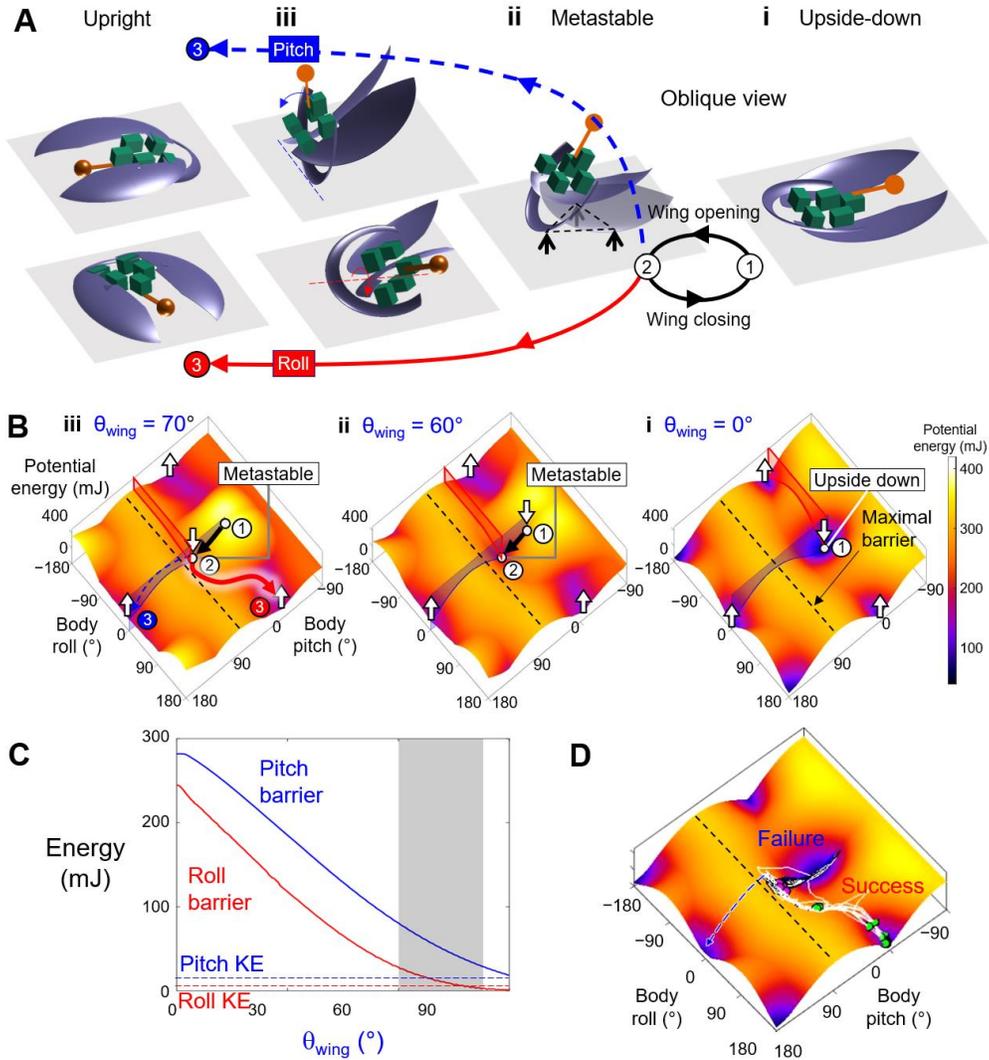

**Fig. 11. Refined robot's self-righting motion and evolving potential energy landscape. (A)** From right to left: schematic of robophysical model in (i) upside-down and (ii) metastable states, and (iii) self-righting by pure pitching (top) and pitching then rolling (bottom). The top and bottom cases correspond to the observed successful and failed attempts in Fig. 9B. **(B)** Corresponding evolving potential energy landscape at different wing opening angles $\theta_{wing}$. 1-3 show upside-down (1), metastable (2), and upright (3) states. Translucent blue and red cross-sectional cuts through the upside-down or metastable local minimum along the pitch and roll axes are used to quantify pitch and roll potential energy barrier. **(C)** Pitch (blue) and roll (red) potential energy barrier as a function of wing opening angle $\theta_{wing}$. Blue and red dashed lines show average maximal pitch and roll kinetic energy, respectively. Gray band shows range of wing opening

amplitudes $\Theta_{wing}$ tested for this robophysical model. **(D)** Ensemble of system state trajectories from all the trials with a given wing opening magnitude. Black: failure to self-right by pure pitching, but being attracted to and trapped in metastable basin. White: successfully self-righting by pitching then rolling. Adapted from [41].

The potential energy landscape changes with wing angle $\theta_{wing}$ as the wings open and close (Fig. 11A, B). To self-right, the system has to escape from a metastable stability basin on the evolving potential energy landscape (Fig. 11B, ii, iii, white dot) and cross a potential energy barrier to reach one of several possible upright basins (Fig. 11B, upward white arrows). When the wings are closed (Fig. 11A, i), the robot is trapped in an upside-down basin (Fig. 11B, i, downward white arrow). As the wings open (Fig. 11A, ii), the upside-down basin shrinks to a metastable basin (Fig. 11B, ii, black arrow), which corresponds to the metastable state with a triangular base of support (Fig. 10B, black dashed triangle). As the wings continue to open (Fig. 11A, iii), the metastable basin becomes higher and moves closer (Fig. 11B, iii, black arrow) to the maximal potential energy barrier (Fig. 11B, black dashed line), which occurs for pure body pitching.

The increasing height of the metastable basin effectively reduces the pitch barrier (Fig. 11C, solid blue; pitch barrier is measured using the blue translucent planes in Fig. 11B) that can be overcome probabilistically by kinetic energy gained along the pitch direction when wing opening stops. However, because the refined robot's self-righting is strenuous by design, even at the maximal wing opening tested, the robot cannot gain enough kinetic energy along the pitch direction to cross the pitch barrier (Fig. 11C, blue, dashed vs. solid in the gray band) to reach the pitch upright basin (Fig. 11B, upward white arrow labeled by blue circle 3), i.e., it cannot self-right by pure pitching (Fig. 11A, top leftmost). Thus, it is trapped in metastable basin, when there are no leg oscillations to inject kinetic energy along the roll direction (Fig. 11D, black, failure trajectories).

However, wing opening also reduces the barrier along the roll direction (Fig. 11C, solid red; roll barrier is measured using the red translucent planes in Fig. 11B). This allows the small kinetic energy along the roll direction from perturbing leg oscillations to overcome the roll barrier probabilistically (Fig. 11C, red, dashed vs. solid in the gray band). Thus, the robot can reach the upright roll basin (Fig. 11B, iii, red arrow, reaching the upward white arrow labeled by red circle 3), i.e., it self-rights by rolling after pitching (Fig. 11A, bottom leftmost), when there are sufficient leg oscillations (Fig. 11D, white, success trajectories). As a result, the larger the robot leg oscillation makes self-righting more probable and reduces the number of attempts required.

Besides flailing legs that we modeled, other perturbing motions observed in the animal likely also contribute to self-righting. For example, small forces from legs scraping the ground [40] may also inject roll kinetic energy. Abdominal flexion and twisting and passive wing deformation under load [40] should tilt the potential energy landscape towards one side and lower the roll barrier on that side. Both these effects should make self-righting easier.

**Stereotyped body rotation results from physical constraints**

When the reconstructed robot rotation trajectories are visualized on the landscape, it is clear that the physical interaction with the ground, which is modeled by a stochastic, self-propelled system with kinetic energy moving on the potential energy landscape, strongly constrains the stochastic system's behavior, resulting in a stereotyped ensemble of trajectories for both successful and failed attempts (Fig. 11D) [41]. This suggested that the discoid cockroach's stereotyped body rotation is largely a result of the physical constraint, consistent with previous findings that physical constraints lead to stereotyped legged locomotion on level, flat, solid ground [54].

**Robot simulation to study effect of substantial randomness in appendage coordination**

Next, we studied the third question—whether the substantial randomness in the coordination of the wing and leg motions is beneficial. To do so, we first created a multi-body dynamics simulation of the refined robot (Fig. 12A) [42]. After being validated against the physical robot experiments (Fig. 10E, see more detail in [42]), the robot simulation allowed us to control and vary the level of motion randomness systematically and collect a large number of trials required to understand its impact, which is less practical in the physical robot and impossible in the animal.

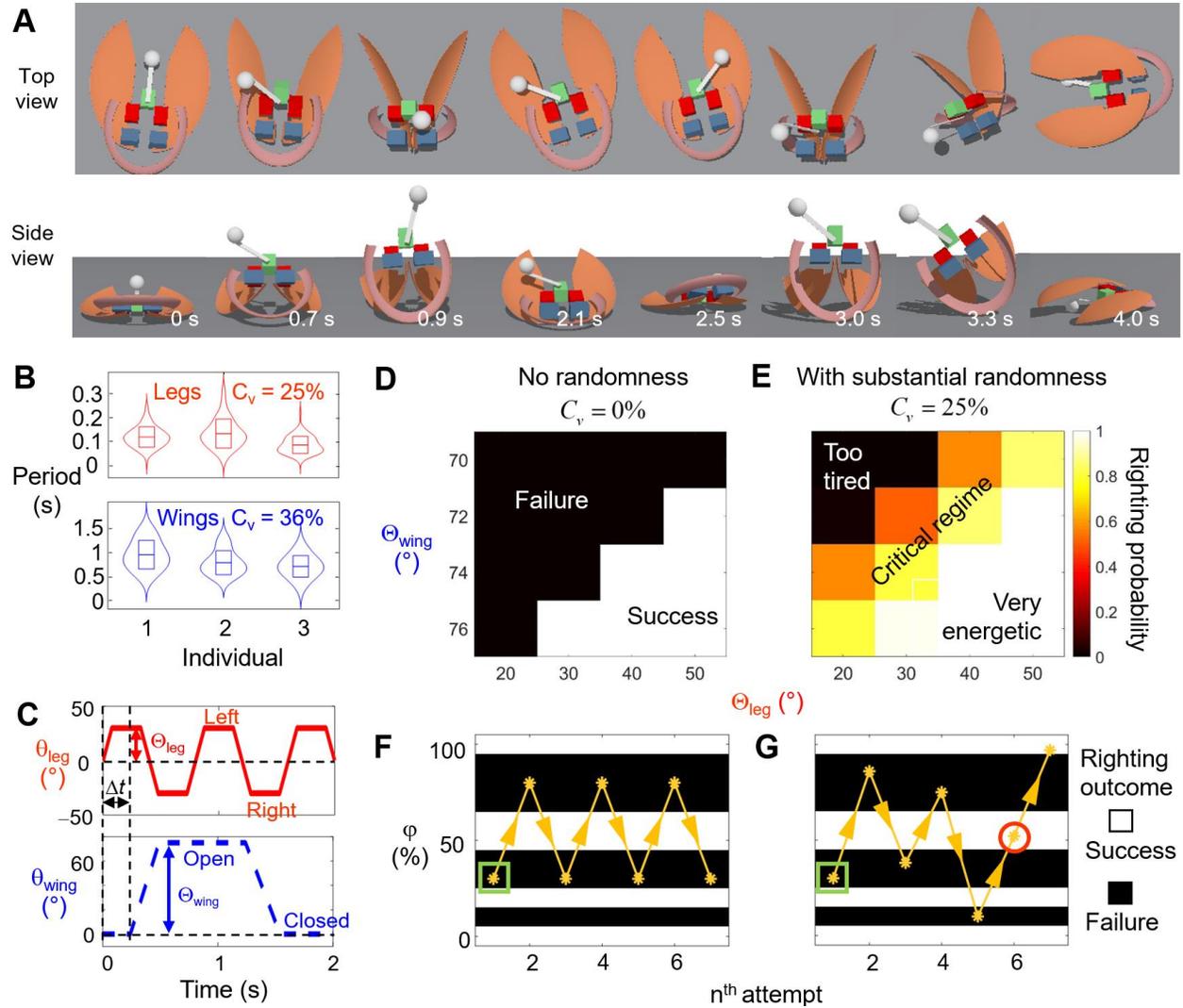

**Fig. 12. Robot simulation reveals benefit of substantial randomness in appendage coordination. (A)** Representative snapshots of simulation robot self-righting after two attempts. Note the resemblance to the refined robot experiments in Fig. 10E. **(B)** Violin plots of wing opening/closing and leg oscillation periods for three discoid cockroach individuals. Inner rectangle shows mean ± 1 s.d. The level of randomness is measured by coefficient of variation, $C_v$ = s.d/mean. **(C)** Actuation profiles of wings (blue) and leg (red) angles (see definition in Fig. 10C, D) of simulation robot. $\Theta_{wing}$ and $\Theta_{leg}$ are wing opening amplitude and leg oscillation amplitude, respectively. $\Delta t$ is the time delay, defined as the time interval between the start

of wing opening and the start of the preceding leg oscillation. Gaussian noise δ*t* is added to Δ*t* in simulation to introduce randomness in phase offset between wing and leg oscillations for each cycle, which can be varied to change the overall randomness level (measured by $C_v$). **(D, E)** Self-righting probability of the simulated robot as a function of $\Theta_{wing}$ and $\Theta_{leg}$, comparing without ($C_v = 0$) and with substantial ($C_v = 25\%$) randomness in wing-leg coordination (phase offset). **(F, G)** Evolution of phase offset φ over consecutive attempts (yellow points connected by arrows), overlaid on phase offset map (white: good phase offsets resulting in successful self-righting, black: bad phase offsets leading to failure), comparing without ($C_v = 0$) and with substantial ($C_v = 25\%$) randomness. Green box shows initial phase offset at the first attempt. Red circle shows the first good phase offset reached with substantial randomness, resulting in success. Adapted from [42].

**Substantial randomness in appendage coordination increases self-righting probability**

Our simulation study revealed that a substantial level of randomness in the motions allows the system to find good coordination between propelling and perturbing appendages, which is more likely to lead to successful self-righting [42]. We measured the levels of randomness in the discoid cockroach's wing opening/closing and leg oscillation periods (Fig. 12B). They are substantially higher than that of the highly rhythmic leg oscillations during walking [51] and running [52]. This substantial level of randomness in both wing and leg oscillation periods results in substantial level of randomness in the phase offset between these two oscillations. Thus, to study the effect of randomness, in the robot simulation, we added Gaussian noise to the time delay between wing and leg oscillations (see definition in Fig. 12C), which injects a similar level of randomness to the coordination (phase offset) between them. When wing opening magnitude $\Theta_{wing}$ and leg oscillation magnitude $\Theta_{leg}$ are small (representing an animal that is too tired), self-righting always fails, whether there is substantial randomness or not (Fig. 12D, E, top left regions). When $\Theta_{wing}$ and $\Theta_{leg}$ are large (representing an animal that is very energetic), self-righting always succeeds, whether there is substantial randomness or not (Fig. 12D, E, bottom right regions). However, for intermediate $\Theta_{wing}$ and $\Theta_{leg}$, when the robot can nearly overcome the potential energy barrier, the substantial randomness increases self-righting probability from 0 to > 40% (Fig. 12D vs. E, critical regime). Because ground self-righting is so strenuous that the animal often barely overcomes the barrier, this finding suggested that the substantial randomness observed in the animals is beneficial to them.

**Substantial randomness helps find good coordination that leads to success**

Further simulation revealed why substantial randomness in the coordination between propelling wings and perturbing legs increases self-righting probability [42]. The leg-wing phase offset has a direct impact on self-righting outcome: good phase offsets almost always lead to success (Fig. 12F, G, white), whereas bad phase offsets almost always lead to failure (Fig. 12F, G, black). Thus, a substantial level of randomness in phase offset allows the system to explore various phase offsets, thereby increasing the chance of finding a good coordination between them that lead to successful self-righting (Fig. 12G, yellow arrows), whereas strictly periodic motions with no randomness traps the system in bad phase offsets always resulting in failure (Fig. 12F, yellow arrows).

**Template to understand why appendage coordination affects self-righting outcome**

Finally, to understand why a substantial level of randomness is useful, we created a template model of strenuous, leg-assisted winged self-righting (Fig. 13B) [43]. A template is the simplest analytical model, comprised of the fewest components and degrees of freedom, that captures fundamental dynamics of this self-righting behavior [57]. Because successful winged self-righting almost always occurs eventually via rolling in both the discoid cockroach (Fig. 9B) and the refined robot (Fig. 10E, Fig. 11D, Fig. 12A), our 2-D template models the planar rolling dynamics of self-righting and trims away the complexity of the system (from Fig. 13A to Fig. 13B). This allowed writing down closed form equations of motion that can be solved numerically to calculate the dynamics of the system.

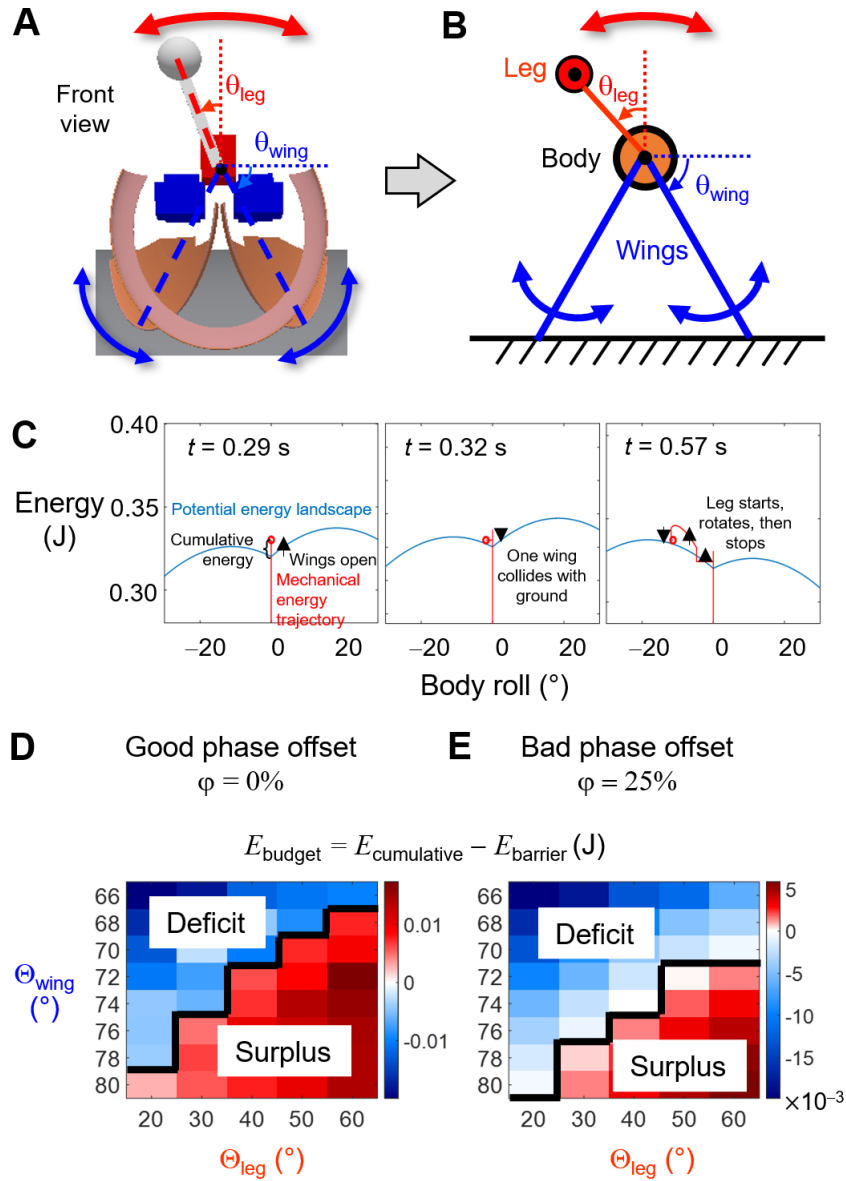

**Fig. 13. Template model reveals why appendage coordination is important.** (**A**) Front view of simulation robot in metastable state. (**B**) Template model capturing planar dynamics of the refined robot. Two point masses represent body (orange) and leg pendulum mass (red). Three massless links represent wings (blue segments) and leg linkage (red segment). In (A, B), blue and red arrows show wing opening/closing and leg oscillation, respectively. (**C**) Example snapshots of system cumulative mechanical energy (red dot) evolution (red trajectory) during a modeling trial, overlaid on evolving potential energy landscape over body roll (blue curve), calculated from the template model. Black bracket defines

cumulative energy, the extra mechanical energy above that of the local minimum of the basin that the system is in. Left: As the wings open. Middle: As one wing collides with the ground. Right: As the leg starts, rotates, and stops. Small upward and downward arrows show cumulative energy increase and reduction in these processes. **(D, E)** Mechanical energy budget as a function of $\Theta_{wing}$ and $\Theta_{leg}$, calculated from the template, comparing without ($C_v = 0$) and with substantial ($C_v = 25\%$) randomness in wing-leg coordination (phase offset). Mechanical energy budget is cumulative energy minus the potential energy barrier: $E_{budget} = E_{cumulative} - E_{barrier}$. The black boundaries separate two regions. The surplus region is where cumulative mechanical energy exceeds the potential energy barrier, leading to successful self-righting. The deficit region is where cumulative mechanical energy is insufficient to overcome the potential energy barrier, resulting in failure. Adapted from [43].

**Good appendage coordination accumulates more energy to overcome barrier**

Successful self-righting requires cumulating sufficient mechanical energy (potential energy and kinetic energy) to overcome the potential energy barrier (which is not always fixed but can be lowered). Thus, to further understand why phase offset affects self-righting outcomes, we used the template model to calculate the system's cumulative energy and barrier, compare whether there is sufficient cumulative energy to overcome the barrier (i.e., mechanical energy budget), and assess how phase offset affects this mechanical energy budget [43].

We calculated the evolving potential energy landscape (potential energy as a function of body roll) of the template model (e.g., Fig. 13C, blue) to obtain the potential energy barrier $E_{barrier}$. We calculated how the system's mechanical energy changes during self-righting attempts (e.g., Fig. 13C, red dot and trajectory) to obtain cumulative energy $E_{cumulative}$, defined as the system's extra mechanical energy above the potential energy of the local minimum of the basin that the system is in (Fig. 13C, left, black bracket). Cumulative energy changes over each actuation phase or collision event: it increases or decreases as the wings open or close (e.g., Fig. 13C, left), increase or decrease as the legs oscillate (e.g., Fig. 13C, right), and always decrease as the robot collides with the ground (e.g., Fig. 13C, middle).

We then used the template model to assess mechanical energy budget, $E_{\text{cumulative}} - E_{\text{barrier}}$, i.e., whether cumulative energy exceeds the potential energy barrier, over a broad range of wing opening and leg oscillation amplitudes, comparing between good and bad phase offsets (e.g., Fig. 13D vs. E) [43]. For both good and bad phase offsets, there is an increasing energy budget surplus as $\Theta_{\text{wing}}$ and $\Theta_{\text{leg}}$ increase (Fig. 13D, E, red regions), and there is an increasing energy deficit as $\Theta_{\text{wing}}$ and $\Theta_{\text{leg}}$ decrease (Fig. 13D, E, blue regions). However, phase offset strongly affects self-righting outcomes by changing mechanical energy budget. Well-coordinated appendage motions with good phase offsets accumulate more mechanical energy than poorly-coordinated ones with bad phase offsets, thereby more effectively overcoming the potential energy barrier (i.e., having a larger energy surplus), and thus self-righting more successfully. Together with the insight that substantial randomness helps find good phase offsets (Fig. 12F, G), this explained why higher randomness increases self-righting probability (Fig. 12D vs. E).

**Summary**

We performed the first studies of biological ground self-righting in three dimensions, using three species of cockroaches. For all three species, ground self-righting is strenuous and may require multiple attempts to succeed (Fig. 4). Two of the three species often self-right dynamically, by generating substantial pitch and/or roll kinetic energy to overcome the potential energy barrier. Each species uses multiple strategies and displays stochastic yet stereotyped transitions across them (Fig. 5, 6). The propelling motion from primary appendages is often accompanied by perturbing motions from other appendages, and all these motions have substantial randomness. Body rotations are complex yet stereotyped (Fig. 8). Compared to failed attempts, in successful attempts their body roll more, which lowers the potential energy barrier (Fig. 8).

We combined robophysical, simulation, and template modeling to understand the physical principles of ground self-righting. Our experiments using an initial robot as a robophysical model revealed that, when propelling against the ground to generate simple planar rotation, the longer and faster appendages push, the more mechanical energy can be gained to overcome the barrier, and thus the more likely and faster self-righting is (Fig. 3). However, the animals can rarely achieve this, because of how strenuous self-righting is for them. To understand the physical principles of strenuous self-righting, we further studied the discoid cockroach's leg-assisted winged self-righting as a model system (Fig. 9). Our robophysical modeling using a refined robot (Fig. 10) revealed that propelling (e.g., wings) or perturbing (e.g., legs) appendages alone cannot gain enough kinetic energy to overcome the high potential energy barrier (Fig. 11). However, when used together, the propelling motion reduces the barrier sufficiently so that it can be overcome probabilistically by the small kinetic energy from perturbing motion (Fig. 11). Thus, only by combining propelling and perturbing motions can self-righting be achieved, when it is so strenuous; this physical constraint (Fig. 11D) leads to the stereotyped body rotation (Fig. 9B). Our robot simulation and template modeling revealed that the substantial randomness observed in the propelling and perturbing motions helps find good coordination between them (Fig. 12), which accumulates more mechanical energy to overcome the potential energy barrier (Fig. 13), thus increasing the likelihood of self-righting.

**Future work**

Further experiments using more species and more elaborate ("anchor"-level [57]) models that better capture the biological detail are needed to generalize the physical principles to diverse biological morphologies and behavior (as well as diverse robot design and actuation). In particular, how limbless and elongate animals [58] self-right on the ground (and even in the air), and how similar robots should do so, remain to be explored. The use of distributed force plates [59] to measure ground reaction forces generated by multiple appendages and body deformation will facilitate this progress. In addition, our findings suggest that animals may use sensory feedback to actively adjust their strategy and appendage motions and body deformation to better self-right, which should be tested by future neurophysiological studies.

Given our progress, the physical principles of ground self-righting in complex terrains remain poorly understood. Natural terrains are rarely perfectly level, flat, solid, and with low friction throughout, can be flowable [60], and can have random objects to grasp onto. Recent animal studies have begun to observe and quantify ground self-righting behavior on surfaces of various roughness [14,15,19,20] and unevenness [14,15], that are flowable [19], or with objects nearby [14]. Rougher or uneven surfaces and random nearby objects facilitate self-righting, and animals adjust their use of diverse strategies correspondingly [14,15,19,20]. In light of these studies, the level, flat, solid, low friction ground used in our studies is likely among the challenging surfaces to self-right on. (It is noteworthy that, despite how strenuous ground self-righting is as our modeling revealed, all three species studied can almost always self-right with 30 seconds (Fig. 4A, white), underscoring the notion that ground self-righting is a crucial locomotor ability that almost all terrestrial animals must possess to survive.) Rougher surfaces likely allow animals to generate larger forces to pitch and/or roll the body to self-right. Similarly, uneven terrain may have asperities of the right sizes [1] for appendages to interlock or even grasp onto to generate large self-righting forces and torques [14]. A sloped surface is presumably easier to self-right on, as animals may rotate on it to gain kinetic energy or slide down to encounter more favorable terrain features. Furthermore, the largely unsuccessful strategies or motions found here may be useful in complex terrains. For example, body twisting may allow legs to reach and grasp onto nearby objects, leg scraping may help engage asperities, and body yawing and sliding may help reach rougher and uneven parts of the terrain, all contributing to self-righting. However, due to their complex mechanics, it is unclear whether flowable substrates make self-righting more or less difficult without modeling the substrate forces [60]. Similarly, in for marine animals on the bottom substrates [22–26], how hydrostatic and hydrodynamic forces work together with substrate forces to achieve underwater ground self-righting is unknown. Future work should measure and model ground self-righting in more complex terrains in three dimensions to elucidate broader physical principles. Our quantitative experimental and modeling approaches demonstrated here will facilitate this progress.

**Acknowledgements**

I would like to thank my students Ratan Othayoth, Qihan Xuan, Toni Wöhrl, Han Lam, and Austin Young and collaborator Chad Kessens, who have worked with me to perform much of the experiments and modeling reviewed here. I am especially grateful to my postdoc advisor, Bob Full, for his visionary guidance from his broad knowledge and strong encouragement over the years that have helped shape my independent scientific journey—much of my lab's work has originated from my time spent in his lab. I am also grateful to my postdoc co-advisor Ron Fearing, for exposing me to his robotics lab, which helped me learn to build robots to use as physical models in my research. Both Bob and Ron's deep appreciation of the usefulness of fundamental principles for robotics helped further solidify my interests in the type of work I am most passionate about. Also thanks to many students for help with preliminary work and animal care, many colleagues for helpful discussion and assistance, and many anonymous reviewers who have helped improve the quality of the research papers reviewed; these individuals are specified in those papers. Three anonymous reviewers provided excellent feedback that greatly improved this review. Thanks to Tonia Hsieh and Marie Schwaner for organizing and inviting me to present at the Symposium on Computational and Physical Models in Research and Teaching to Explore Form-Function Relationships held at the Society for Integrative & Comparative Biology 2024 Annual Meeting.

**Funding**

I was supported by the Burroughs Wellcome Fund Career Award at the Scientific Interface for support for writing this review article. Thanks to National Science Foundation, Company of Biologists, and Society for Integrative & Comparative Biology for supporting my travel to present at the symposium at SICB 2024.

**Conflict of Interest**

I have no conflict of interest.